\documentclass[10pt, conference]{IEEEtran}

\usepackage{amsmath, amssymb}
\usepackage{graphicx}
\usepackage[sans]{dsfont}
\usepackage{stmaryrd}
\usepackage{mathrsfs}
\usepackage{subfigure}
\usepackage{color}
\usepackage{cite}
\usepackage{asymptote}

\DeclareMathAlphabet{\mpzc}{OT1}{pzc}{m}{it}

\newcommand{\msf}{\mathsf}

\newcommand{\mbb}{\mathbb}
\newcommand{\mrm}{\mathrm}
\newcommand{\mcl}{\mathcal}

\newcommand{\mds}{\mathds}

\newcommand\ind{\protect\mathpalette{\protect\independent}{\perp}}
\def\independent#1#2{\mathrel{\rlap{$#1#2$}\mkern2mu{#1#2}}}

\def\p#1{p_{\mathsf{#1}}}

\newtheorem{thm}{Theorem}
\newtheorem{lem}[thm]{Lemma}
\newtheorem{prop}[thm]{Proposition}

\newtheorem{eg}[thm]{Example}
\newtheorem{rem}{Remark}

\title{Modeling and Information Rates for Synchronization Error Channels}
\author{
\authorblockN{Aravind R. Iyengar, Paul H. Siegel, and Jack K. Wolf}
\authorblockA{University of California San Diego, La Jolla, CA $92093-0401$, USA \\Email: \{aravind, psiegel, jwolf\}@ucsd.edu}
}
\newcommand{\twobibs}[2]{#2}
\newcommand{\figselect}[2]{#2}

\begin{document}
\maketitle

\begin{abstract}
We propose a new channel model for channels with synchronization errors. Using this model, we give simple, non-trivial and, in some cases, tight lower bounds on the capacity for certain synchronization error channels.
\end{abstract}

\section{Introduction}

Channels with synchronization errors have been of interest from the very beginnings of information theory. However, little is known of their capacities or of good coding schemes. In the last decade, a flurry of activity has led to significant progress in estimating achievable information rates for certain synchronization error channels (SECs). A ``good'' coding scheme continues to be elusive.

In this paper, we model an SEC as a channel with states and use this model to arrive at some simple lower bounds on the capacity. Although the idea behind the alternative model is straightforward, the model itself has been absent in literature. While the present paper deals only with a few asymptotic results on information rates of the SEC, we think that the model presented here can be utilized to design codes for SECs in general. 

The remainder of this paper is organized as follows. In Section \ref{sec_sec}, we recall a few of the main results on capacity of SECs. We consider a special case of the generic SEC---the deletion, duplication channel (DDC)---and construct an equivalent channel by viewing the SEC as a channel with states in Section \ref{sec_eqchn}. We use the model to obtain bounds on the capacity in Section \ref{sec_bounds}. We conclude by highlighting the possible advantages of the model discussed in Section \ref{sec_conc}.

\section{Synchronization Error Channels} \label{sec_sec}
\begin{rem}[Notation]
Non-random variables are written as lowercase letters, e.g. $n$. We denote sets by double-stroke uppercase letters, $\mbb{X}$, and define $[n] = \{1, 2, \cdots, n\}$, $[0] = \emptyset$ and $[m:n] = \{m, m + 1, \cdots, n\}, m \leq n$. We assume an underlying probability space $(\Omega, \mathcal{F}, \msf{P})$ over which all random variables, denoted by uppercase letters $X$, are defined. Random vectors are denoted by uppercase letters with the set of indices as subscripts, e.g. $X_{[n]} = (X_1, X_2, \cdots, X_n)$ or $X_{Y_{[n]}}$ when the set of indices is itself a random vector $Y_{[n]}$. Random processes are denoted by script letters $\mcl{X}$, or as $X_{\mbb{N}}$.
\end{rem}

Let $\mbb{X}$ be a finite set. A \emph{memoryless} synchronization error channel is specified by a stochastic matrix $\{q(\overline{y}|x), \overline{y} \in \overline{\mbb{Y}}, x \in \mbb{X}\}$ where $\mbb{Y}$ is the output alphabet and $\overline{\mbb{Y}}$ is the set of all strings (including the empty string $\lambda$) over $\mbb{Y}$. We assume that the expected length of the output string arising from one input symbol is strictly positive and finite, i.e. $0 < \sum_{\overline{y} \in \overline{\mbb{Y}}} |\overline{y}|q(\overline{y} | x) < \infty$, where $|\overline{y}|$ denotes the length of the string $\overline{y}$ (the number of symbols in $\overline{y}$). For $x_{[n]} = (x_1, x_2, \cdots, x_n) \in \mbb{X}^n$ and $\overline{y}_{[n]} = (\overline{y}_1, \overline{y}_2, \cdots, \overline{y}_n) \in \overline{\mbb{Y}}^n$, we write $q_n(\overline{y}_{[n]} | x_{[n]}) = \prod_{i = 1}^n q(\overline{y}_i | x_i)$. Let $\overline{\overline{y}_{[n]}}$ denote the concatenation of strings $\overline{y}_i, i \in [n]$. Then a memoryless SEC $Q_n$ is specified by the input alphabet $\mbb{X}$, output alphabet $\mbb{Y}$ and transition probabilities
\begin{equation} \label{eq_msectp}
Q_n(\overline{y} | x_{[n]}) = \sum_{\overline{\overline{y}_{[n]}} = \overline{y}} q_n(\overline{y}_{[n]} | x_{[n]})
\end{equation}
for $\overline{y} \in \overline{\mbb{Y}}$ and $x_{[n]} \in \mbb{X}^n$. Consider the sequence of channels $\{Q_n\}_{n = 1}^{\infty}$ where $Q_n$ is as defined above. Then, the following was shown by Dobrushin in 1967.
\begin{thm}[Coding Theorem \cite{dob_67_pit_sec}] \label{thm_cap}
Let $X_{[n]}$ and $\overline{Y}$ denote the input and the output of the SEC $Q_n$. Let
\[
C_n = \sup_{\mathsf{P}(X_{[n]})} \frac{1}{n}I(X_{[n]}; \overline{Y}).
\]
Then $C = \lim_{n \rightarrow \infty} C_n = \inf_{n \geq 1} C_n$ exists and is equal to the transmission capacity of the SEC. Furthermore,
\[
C = \sup_{\mcl{X}_\mcl{M}} \lim_{n \rightarrow \infty} \frac{1}{n}I(X_{[n]}; \overline{Y})
\]
where $\mcl{X}_\mcl{M}$ is a stationary, ergodic, Markov process over $\mbb{X}$.
\end{thm}

\begin{figure*}[t]
\begin{align}
P_n(\overline{y} | x_{[n]}, Z_0 = 0) &= \sum_{\{\overline{z} : |\overline{z}| = |\overline{y}|\}} \msf{P}(\overline{Z} = \overline{z}, \overline{Y} = \overline{y} | X_{[n]} = x_{[n]}, Z_0 = 0) \notag \\
&= \sum_{\{\overline{z} : |\overline{z}| = |\overline{y}|\}} \msf{P}(\overline{Z} = \overline{z} | Z_0 = 0)\msf{P}(\overline{Y} = \overline{y} | X_{[n]} = x_{[n]}, Z_0 = 0, \overline{Z} = \overline{z}) \notag \\
&= \sum_{\{\overline{z} : |\overline{z}| = |\overline{y}|\}} \prod_{i = 1}^{|\overline{y}|} \Big(\msf{P}(Z_i = z_i | Z_{i - 1} = z_{i - 1}, Z_0 = 0)\msf{P}(Y_i = y_i | X_{[n]} = x_{[n]}, Z_i = z_i)\Big) \label{eq_eqchntrnp}
\end{align}
\hrule
\end{figure*}
We will consider an example of an SEC and confine our attention to this channel throughout this paper.

\begin{eg}[Deletion-Duplication Channel (DDC)] \label{eg_ddc}
Consider the binary SEC with $\mbb{X} = \mbb{Y} = \{0, 1\}$ and the following stochastic matrix
\[
q(\overline{y} | x) =
\begin{cases}
p_{\msf{d}}, &\overline{y} = \lambda\\
p_{\msf{t}}p_{\msf{i}}^{r - 1}, &\overline{y} = x^r,{\ }\forall{\ }r \geq 1
\end{cases}
\]
with $p_{\msf{d}} + \sum_{r = 1}^{\infty} p_\msf{t}p_\msf{i}^{r - 1} = 1 \Rightarrow p_\msf{t} = (1 - p_\msf{d})(1 - p_\msf{i})$ for $p_\msf{i} < 1$. This model implies that deletions and duplications occur i.i.d. (and mutually exclusively) with probabilities $p_\msf{d}$ and $p_\msf{i}$ respectively. The expected output string length is
\[
0 < \sum_{r = 1}^\infty rp_\msf{t}p_\msf{i}^{r - 1} = \frac{p_\msf{t}}{(1 - p_\msf{i})^2} = \frac{1 - p_\msf{d}}{1 - p_\msf{i}} < \infty.
\]
Hence $(p_\msf{i}, p_\msf{d}) \in [0, 1)^2$. Since the capacity is zero when either $\p{i}$ or $\p{d}$ is $1$ (they cannot simultaneously be $1$), this model does indeed represent the entire class of deletion-duplication channels. Note that when $\p{i} = 0$, the DDC is the same as the \emph{binary deletion channel} (BDC), and when $\p{d} = 0$, it is the so-called \emph{binary sticky channel} \cite{mit_08_tit_sticky}.
\end{eg}

The BDC has been the most well-studied SEC. In \cite{mit_09_prs_delchn}, the author surveys the results that were known prior to 2009. To summarize, the best known lower bounds were obtained, chronologically, through bounds on the cutoff rate for sequential decoding \cite{gal_61_llr_seqdecsec}, bounding the rate with a first-order Markov input \cite{dig_06_tit_finbufchn}, reduction to a Poisson-repeat channel \cite{mit_06_tit_lbcapdc}, analyzing a ``jigsaw-puzzle'' coding scheme \cite{dri_07_tit_implbddc}, or by directly bounding the information rate by analyzing the channel as a joint renewal process \cite{kir_10_tit_ddccap}. Recently, \cite{kan_10_isit_delchn} and \cite{kal_10_isit_delchn} independently gave the capacity of a BDC with small deletion probabilities, and showed that it is achieved by independent and uniformly distributed (i.u.d.) inputs. The known upper bounds for the BDC have been obtained by genie-aided decoder arguments \cite{dig_07_isit_capubdc, fer_10_tit_bdccap}. An idea from \cite{fer_10_tit_bdccap} was extended to obtain some analytical lower bounds on the capacity of channels that involve substitution errors as well as insertions or deletions \cite{rah_11_arx_capid}. In contrast to these existing results, our approach explicitly characterizes the achievable information rates in terms of ``subsequence-weights'', which is a measure relevant in ML decoding for the BDC \cite{mit_09_prs_delchn}. Additionally, the method proposed here gives the tight bound on capacity for small deletion probabilities obtained in \cite{kan_10_isit_delchn} more directly.

For the sticky channel, \cite{mit_08_tit_sticky} obtained lower bounds on the capacity by numerically estimating the capacity per unit cost of the equivalent channel of runs through optimization of $8$ and $16$ bit codes. Here, we obtain direct analytical lower bounds on the capacity. These, to the best of our knowledge, represent the only analytical bounds for the capacity of the sticky channel.

\section{SEC as a Channel with States} \label{sec_eqchn}
For the DDC $Q_n$, we write
\begin{equation} \label{eq_eqchnmod}
Y_i = X_{\Gamma_i} = X_{i - Z_i}, i \in [N_n]
\end{equation}
where $Z_i \in \mbb{Z}$ is the ``state'' of the channel and we define the length of the output to be $N_n \triangleq \sup \{i \geq 0 : \Gamma_i \leq n | \Gamma_0 = 0\}$. The state process $\mcl{Z}$ is independent of the channel input process $\mathcal{X}$, and is a \emph{time-homogeneous} Markov chain over the set of integers $\mbb{Z}$ with transition probabilities
\begin{equation} \label{eq_ztp}
\msf{P}(Z_i = z_i | Z_{i - 1} = z_{i - 1}) =
\begin{cases}
\p{i}, &z_i = z_{i - 1} + 1\\
\p{t}\p{d}^r, &z_i = z_{i - 1} - r, r \geq 0
\end{cases}
\end{equation}
where we define $\p{t} = (1 - \p{d})(1 - \p{i})$ for normalization. We also assume the boundary condition that $Z_0 = 0$, i.e., that there was perfect synchronization initially. It is easy to see that $N_n < \infty{\ }\forall{\ }n \in \mbb{N}$ a.s. since we impose $\p{i} < 1$, and that $N_n \rightarrow \infty$ a.s. as $n \rightarrow \infty$ since $\p{d} < 1$. We refer to the $\Gamma$-process as the \emph{index process}. The index process and the channel state process have a one-to-one correspondence, and consequently, we will use them interchangeably. From the state transition probabilities in \eqref{eq_ztp}, it is also clear that the $\mcl{Z}$ process is \emph{shift-invariant}, i.e., $\msf{P}(Z_i = z_i | Z_{i - 1} = z_{i - 1}) = \msf{P}(Z_i = z_i -  z_{i - 1} | Z_{i - 1} = 0)$. The index process inherits this property from the $\mcl{Z}$ process.

For $\overline{y} \in \overline{\mbb{Y}}$ and $x_{[n]} \in \mbb{X}^n$, the channel transition probabilities are given as in Equation \eqref{eq_eqchntrnp}. Note that in the terms within the parenthesis on the right hand side of Equation \eqref{eq_eqchntrnp}, the first term is completely specified by the transition probabilities \eqref{eq_ztp} of the channel state process $\mcl{Z}$, and the second term is $0$ or $1$ accordingly as $y_i = x_{i - z_i}$ or $y_i \neq x_{i - z_i}$ respectively. The input and output alphabets of $P_n$ are $\mbb{X} = \mbb{Y} = \{0, 1\}$. The equivalence between the DDC $Q_n$ and $P_n$, for any $n$, is evident by noting that for every parsing of $\overline{y} \in \overline{\mbb{Y}}$ as $\overline{y}_{[n]}$ in Equation \eqref{eq_msectp}, there is a corresponding state path $\overline{z} \in \overline{\mbb{Z}}$ in Equation \eqref{eq_eqchntrnp} (and vice versa) and that the terms within the parenthesis in \eqref{eq_eqchntrnp}, when grouped according to the output symbols arising from the same input symbol, spell out exactly the same probability as the terms $q(\overline
 {y}_i | x_i)$.

As a consequence of the above equivalence, Theorem \ref{thm_cap} applies to the sequence of channels $\{P_n\}_{n = 1}^{\infty}$ specified by Equations \eqref{eq_eqchnmod} and \eqref{eq_ztp}. We hence have for input $X_{[n]}$ and output $Y_{[N_n]}$ of $P_n$,
\begin{align}
C &= \lim_{n \rightarrow \infty} \sup_{\msf{P}(X_{[n]})} \frac{1}{n} I(X_{[n]}; Y_{[N_n]}) \notag \\
&= \sup_{\mcl{X}_\mcl{M}} \lim_{n \rightarrow \infty} \frac{1}{n} I(X_{[n]}; Y_{[N_n]}). \notag
\end{align}
We will restrict our attention to stationary, ergodic, Markov sources $\mcl{X}_\mcl{M}$. Under this assumption, the output process $\mcl{Y}$ is also stationary, and the \emph{entropy rate} $\mcl{H}(\mcl{Y})$ is well-defined.

\section{Bounds on Capacity} \label{sec_bounds}
With the setting of Section \ref{sec_eqchn}, it is possible to immediately obtain some non-trivial bounds on the capacity. We start with some simple bounds and bounding techniques for the DDC and consider the BDC and the sticky channel in separate subsections.

\begin{prop} \label{prop_capbounds}
For the deletion-duplication channel,
\[
\Big((1 - \p{d})\Big(1 - \frac{h_2(\p{i})}{1 - \p{i}}\Big) - h_2(\p{d})\Big)^+ \leq C \leq 1 - \p{d},
\]
where $(x)^+ = \max\{0, x\}$ and $h_2(\cdot)$ is the \emph{binary entropy function}.
\end{prop}
\begin{IEEEproof}
We can write
\begin{align}
I(X_{[n]}; &Y_{[N_n]}) = I(X_{[n]}; Y_{[N_n]}, Z_{[N_n]}) - I(X_{[n]}; Z_{[N_n]} | Y_{[N_n]}) \notag \\
&\stackrel{(a)}{=} I(X_{[n]}; Y_{[N_n]} | Z_{[N_n]}) - I(X_{[n]}; Z_{[N_n]} | Y_{[N_n]}) \notag \\
&\stackrel{(b)}{=} (1 - \p{d})H(X_{[n]}) - I(X_{[n]}; Z_{[N_n]} | Y_{[N_n]}), \label{eq_capeq}
\end{align}
where $(a)$ is true because $\mcl{X} \ind \mcl{Z}$ and $(b)$ from the fact that the DDC, given the $\mcl{Z}$ process realization, is equivalent to a BEC with erasure rate $\p{d}$. Then,
\[
n(1 - \p{d}) \geq I(X_{[n]}; Y_{[N_n]}) \geq (1 - \p{d})H(X_{[n]}) - H(Z_{[N_n]}).
\]
Since the $\mcl{Z}$ process is a Markov chain, we can easily show $H(Z_{[N_n]}) \leq \msf{E}(N_n)\Big(h_2(\p{i}) + \frac{1 - \p{i}}{1 - \p{d}}h_2(\p{d})\Big)$, where the inequality follows because, for any finite $n$, we have the extra knowledge that $Z_i \geq i - n$ by definition. This extra knowledge becomes tautological when $n \rightarrow \infty$ so that
\[
\lim_{n \rightarrow \infty} \frac{H(Z_{[N_n]})}{n} = \Big(\lim_{n \rightarrow \infty} \frac{\msf{E}(N_n)}{n}\Big)\Big(h_2(\p{i}) + \frac{1 - \p{i}}{1 - \p{d}}h_2(\p{d})\Big).
\]
Writing the $\Gamma$-process as a renewal process, from the strong law of large numbers, $\frac{N_n}{n} \rightarrow \frac{1 - \p{d}}{1 - \p{i}}$ a.s. as $n \rightarrow \infty$.
\end{IEEEproof}

Note that the above result implies the following for the BDC ($\p{i} = 0, \p{d} = p$), the \emph{symmetric} DDC (with $\p{i} = \p{d} = p$) and the sticky channel ($\p{i} = p, \p{d} = 0$) respectively.
\begin{align}
(1 - p - h_2(p))^+ &\leq C_{\mrm{BDC}} \leq 1 - p, \notag \\
(1 - p - 2h_2(p))^+ &\leq C_{\mrm{SDDC}} \leq 1 - p, \notag \\
\Big(1 - \frac{h_2(p)}{1 - p}\Big)^+ &\leq C_{\mrm{Sticky}} \leq 1. \notag
\end{align}
Although these bounds have simple closed-form expressions, they are far from the best known bounds for the capacity of these channels. We can, however, improve these bounds. We have from Equation \eqref{eq_capeq},
\begin{align}
I(X_{[n]}; Y_{[N_n]}) &= (1 - \p{d})H(X_{[n]}) + I(Y_{[N_n]}; Z_{[N_n]}) \notag \\
&\quad - H(Z_{[N_n]}) + H(Z_{[N_n]} | X_{[n]}, Y_{[N_n]}). \label{eq_ixy}
\end{align}
We can easily show that $\mcl{H}(\mcl{Z} | \mcl{X}, \mcl{Y}) = \frac{1 - \p{d}}{1 - \p{i}}H(Z_1 | \mcl{X}, \mcl{Y})$, and hence
\begin{align}
C &\geq \sup_{\mcl{X}} \Big(\mcl{H}(\mcl{X}) + \frac{H(Z_1 | \mcl{X}, \mcl{Y})}{1 - \p{i}}\Big)(1 - \p{d}) \notag \\
&\hspace*{30mm} -\frac{1 - \p{d}}{1 - \p{i}}h_2(\p{i}) - h_2(\p{d}). \label{eq_capddc}
\end{align}
It is not easy to evaluate the right hand side of the above inequality. However, we can lower bound it further by introducing some conditioning.
\begin{align}
C &\geq \sup_{\mcl{X}} \Big(\mcl{H}(\mcl{X}) + \frac{H(Z_1 | Z_i, \mcl{X}, \mcl{Y})}{1 - \p{i}}\Big)(1 - \p{d}) \notag \\
&\hspace*{10mm} - \frac{1 - \p{d}}{1 - \p{i}}h_2(\p{i}) - h_2(\p{d}) \stackrel{\Delta}{=} \sup_{\mcl{X}} L_i^{\mcl{X}}{\ }\forall{\ }i \in \mbb{N}. \notag
\end{align}
\begin{lem} \label{lem_monli}
The sequence $\{L_i^{\mcl{X}}\}_{i = 1}^{\infty}$ is non-decreasing.
\end{lem}
\begin{IEEEproof}
We have
\begin{align}
H(Z_1 | Z_{i + 1}) &= H(Z_1, Z_i | Z_{i + 1}) - H(Z_i | Z_1, Z_{i + 1}) \notag \\
&\stackrel{(a)}{=} H(Z_i | Z_{i + 1}) + H(Z_1 | Z_i) - H(Z_i | Z_1, Z_{i + 1}) \notag \\
&= H(Z_1 | Z_i) + I(Z_1 ; Z_i | Z_{i + 1}) \geq H(Z_1 | Z_i), \notag
\end{align}
where $(a)$ follows from the Markovity of the $\mcl{Z}$ process. Since conditioning on $\mcl{X}$ and $\mcl{Y}$ retains the above chain of inequalities, we have $H(Z_1 | Z_{i + 1}, \mcl{X}, \mcl{Y}) \geq H(Z_1 | Z_i, \mcl{X}, \mcl{Y}){\ }\forall{\ }i \geq 1$. Hence $\{L_i^{\mcl{X}}\}_{i = 1}^\infty$ is non-decreasing, and maximizing over stationary, ergodic, Markov input processes $\mcl{X}$ gives the bound in Proposition \ref{prop_capbounds} for $i = 1$. Therefore, for increasing $i$, we have bounds better than the one in Proposition \ref{prop_capbounds}, and in the limit as $i \rightarrow \infty$, we approach the bound in \eqref{eq_capddc}.
\end{IEEEproof}
For the case of the BDC and the sticky channel, evaluating some of these bounds is easier, owing to the fact that the $\mcl{Z}$ process is monotonic, i.e., in these cases, the output is just a subsequence of the input sequence and vice versa respectively.

\subsection{Information Rates for the BDC}
\begin{figure*}[!ht]
\begin{align} \label{eq_him}
&\mathfrak{H}^{(i)}_m = \sum_{x_{[m + i - 1]}}\frac{1}{2^{m + i - 1}}\Big(\sum_{y_{[i - 1]}}\frac{w_{y_{[i - 1]}}(x_{[m + i - 1]})}{{m + i - 1 \choose m}}\mathfrak{h}(x_{[m + i - 1]}, y_{[i - 1]})\Big), \\
\mathfrak{h}(x_{[m + i -1]}, y_{[i - 1]}) &= -\sum_{z = -m}^0\mds{1}_{\{x_{1 - z} = y_1\}}\frac{w_{y_{[2 : i - 1]}}(x_{[2 - z : m + i - 1]})}{w_{y_{[i - 1]}}(x_{[m + i - 1]})}\log_2\Big(\mds{1}_{\{x_{1 - z} = y_1\}}\frac{w_{y_{[2 : i - 1]}}(x_{[2 - z : m + i - 1]})}{w_{y_{[i - 1]}}(x_{[m + i - 1]})}\Big). \notag
\end{align}
\hrule
\end{figure*}

For the BDC with i.u.d. inputs, we can easily show that $\mcl{Y}$ is also an i.u.d. sequence. Consequently, $\mcl{I}(\mcl{Y}; \mcl{Z}) = 0$ because the only information obtained from $Y_{[N_n]}$ about $Z_{[N_n]}$ is the length of the vector, and this information vanishes in the limit as $n \rightarrow \infty$ as a result of the concentration. Therefore, we have from Equation \eqref{eq_ixy} that the lower bound in Equation \eqref{eq_capddc} is actually the \emph{symmetric information rate} (SIR) in this case. Let us denote by $w_{y_{[i]}}(x_{[j]})$ the number of subsequences of $x_{[j]}$ that are the same as $y_{[i]}$, and define $w_{\lambda}(x_{[j]}) = 1{\ }\forall{\ }x_{[j]}$. We will refer to $w_{y_{[i]}}(x_{[j]})$ as the \emph{$y_{[i]}$-subsequence weight} of the vector $x_{[j]}$.

We will focus on the term $H(Z_1 | Z_i, \mcl{X}, \mcl{Y})$ which is the only term to be evaluated to get an estimate of $L_i^{\mrm{iud}}$. First note that $H(Z_1 | Z_i, \mcl{X}, \mcl{Y}) = H(Z_1 | Z_i, X_{[i - 1 - Z_i]}, Y_{[i - 1]})$. Given $Z_i = -m, X_{[m + i - 1]} = x_{[m + i - 1]}$ and $Y_{[i - 1]} = y_{[i - 1]}$, we have $Z_1 \in \{j \in \{0, -1, \cdots, -m\} : x_{1 - j} = y_1, w_{y_{[2 : i - 1]}}(x_{[2 - j : m + i - 1]}) > 0\}$. Further, it is easy to see that
\begin{align}
\msf{P}(&Z_1 = z | Z_i = -m, X_{[m + i - 1]} = x_{[m + i - 1]}, Y_{[i - 1]} = y_{[i - 1]}) \notag \\
&= \mds{1}_{\{x_{1 - z} = y_1\}}\frac{w_{y_{[2 : i - 1]}}(x_{[2 - z : m + i - 1]})}{w_{y_{[i - 1]}}(x_{[m + i - 1]})}, -m \leq z \leq 0.\notag
\end{align}
Since $\msf{P}(X_{[m + i - 1]} = x_{[m + i - 1]} | Z_i = -m) = 2^{-(m + i - 1)}$,
\begin{align}
\msf{P}(Y_{[i - 1]} = y_{[i - 1]} &| X_{[m + i - 1]} = x_{[m + i - 1]}, Z_i = -m) \notag \\
&= \frac{w_{y_{[i - 1]}}(x_{[m + i - 1]})}{{m + i - 1 \choose m}}, \notag
\end{align}
and $\msf{P}(Z_i = -m | Z_0 = 0) = \msf{p}^{\otimes i}(-m)$, where we write $\msf{p}(-m) = \msf{P}(Z_1 = -m | Z_0 = 0)$, $\msf{p}^{\otimes i}(-m) = (\msf{p} \otimes \msf{p}^{\otimes i - 1})(-m)$, $\msf{p}^{\otimes 1}(-m) = \msf{p}(-m)$ with $\otimes$ denoting convolution, we have for any $i \in \mbb{N}$
\[
C \geq L_i^{\mrm{iud}} = \Big(1 + \sum_{m \geq 0} \msf{p}^{\otimes i}(-m)\mathfrak{H}^{(i)}_m\Big)(1 - p) - h_2(p)
\]
where $\mathfrak{H}^{(i)}_m$ is as given in Equation \eqref{eq_him}. Unfortunately, evaluating $\mathfrak{H}^{(i)}_m$ for $i > 2$ is hard since counting subsequences is not easy. For the case of $i = 2$, we can easily evaluate
\begin{align}
L_2^{\mrm{iud}} = \Big(1 + (1 - p)^2 \sum_{m \geq 0} (m + 1)p^m\mathfrak{H}_m^{(2)}\Big)(1 - p) - h_2(p) \label{eq_l2sir}
\end{align}
with
\[
\mathfrak{H}_m^{(2)} = \log_2(m + 1) - \frac{1}{2^{m + 1}}\sum_{i = 0}^{m + 1}{m + 1\choose i}h_2\Big(\frac{i}{m + 1}\Big).
\]
Although evaluating $L_i^{\mrm{iud}}$ for $i > 2$ is hard, we can further lower bound it as follows.
\begin{align}
L_i^{\mrm{iud}} &= (1 + H(Z_1 | Z_i, \mcl{X}, \mcl{Y}))(1 - p) - h_2(p) \notag \\
&\geq \Big(1 + \sum_{m = 0}^j \msf{P}(Z_i = -m) H(Z_1 | Z_i = -m, \mcl{X}, \mcl{Y})\Big) \notag \\
&\qquad\qquad\times (1- p) - h_2(p) \notag \\
&\stackrel{\Delta}{=} 1 - p - h_2(p) + (1 - p)a^{(i)}_j \notag \\
&\stackrel{\Delta}{=} \mathfrak{L}^{(i)}_j{\ }\forall{\ }j \geq 0, i \geq 1. \notag
\end{align}
We can then use $C \geq \sup_{i \geq 1}\mathfrak{L}^{(i)}_j \stackrel{\Delta}{=} \mathfrak{L}^{\mrm{iud}}_j$ for some $j \geq 0$ as a lower bound for the capacity. We proceed as follows
\begin{align}
a^{(i)}_j &= a_{j - 1}^{(i)} + \msf{p}^{\otimes i}(-j)\cdot H(Z_1 | Z_i = -j, \mcl{X}, \mcl{Y}) \notag \\
&\stackrel{\Delta}{=} a_{j - 1}^{(i)} + \msf{p}^{\otimes i}(-j)b_j^{(i)}. \label{eq_aterm}
\end{align}
Since $a^{(i)}_0 = b^{(i)}_0 = 0$, we have $a^{(i)}_1 = ip(1 - p)^ib^{(i)}_1$. Let us denote by $r_1(x_{[n]})$ the length of the first run in the vector $x_{[n]}$. Then, we can show that, for $y_{[i - 1]}$ received from $x_{[i]}$ with a single deletion, 
\[
H(Z_1 | Z_i = -1, X_{[i]} = x_{[i]}, Y_{[i - 1]} = y_{[i - 1]}) = h_2\Big(\frac{1}{r_1(x_{[i]})}\Big),
\]
and hence
\begin{align}
b_1^{(i)} &= \sum_{x_{[i]}}\frac{1}{2^i}\sum_{y_{[i - 1]}}\frac{1}{i}h_2\Big(\frac{1}{r_1(x_{[i]})}\Big) 
. \label{eq_aiterm}
\end{align}
From \eqref{eq_aterm} and \eqref{eq_aiterm},
\[
a_1^{(i)} = p(1 - p)^i\sum_{j = 1}^i \frac{j}{2^j}h_2\Big(\frac{1}{j}\Big) + p(1 - p)^i\frac{i}{2^i}h_2\Big(\frac{1}{i}\Big)
\]
and thus
\begin{align}
C &\geq 
1 + p\log_2p - cp + O(p^2) \label{eq_l1sir}
\end{align}
where $c = \log_2(2e) - \frac{1}{2}\sum_{j \geq 1}\frac{j}{2^j}\log_2j \approx 1.154163$. Note that this is exactly the expression obtained for capacity for small $p$ in \cite{kan_10_isit_delchn}. In the evaluation of the above bound, we were helped by the fact that when we restrict to the case of a single deletion, the ambiguity in the first channel state $Z_1$ arises only when $r_1(x_{[i]}) > 1$, in which case the uncertainty is exactly $h_2\Big(\frac{1}{r_1(x_{[i]})}\Big)$. This, however, is not true when there are $2$ or more deletions, wherein we will have to count subsequence weights of sequences.

We can obtain similar bounds for symmetric first-order Markov input processes. 
But these calculations will have to keep track of \emph{ascents} and \emph{descents} in sequences, and are therefore more tedious. We can write for $\msf{P}(X_i = x \oplus 1 | X_{i - 1} = x) = \alpha$,
\begin{align}
L_2^{\mcl{M}1} = \Big[\sup_{\alpha} \Big(h_2(\alpha) + (1 - p)^2\sum_{m \geq 0}(m + 1)p^m\ell_m(\alpha)\Big)\Big] \notag \\
\qquad\qquad\qquad\times (1 - p) - h_2(p) \text{, where} \notag \\
\ell_m(\alpha) = \log_2(m + 1) - \sum_{i = 0}^{m + 1}h_2\Big(\frac{i}{m + 1}\Big)\pi(\alpha, i, m + 1), \notag
\end{align}
and $\pi(\cdots)$ is defined recursively as
\begin{align}
\pi(\alpha, i, m) &= \frac{1}{2}\pi_0(\alpha, i, m) + \frac{1}{2}\pi_1(\alpha, i, m) \notag \\
\pi_0(\alpha, i, m) &= (1 - \alpha)\pi_0(\alpha, i, m - 1) + \alpha\pi_1(\alpha, i - 1, m - 1) \notag \\
\pi_1(\alpha, i, m) &= (1 - \alpha)\pi_1(\alpha, i - 1, m - 1) + \alpha\pi_0(\alpha, i, m - 1) \notag
\end{align}
with $\pi_j(\alpha, i, m) = \frac{1}{2}(1 - \alpha)^{m - 1}, i \in \{0, m\}, j \in \{0, 1\}$ and $\pi_j(\alpha, i, m) = 0$ for $i \notin [m], j \in \{0, 1\}$. We can also evaluate
\begin{align}
&\mathfrak{L}_1^{\mcl{M}1} = -h_2(p) + (1 - p) \times \notag \\
&\sup_{\alpha}\Big[h_2(\alpha) + p\cdot\sup_{i \geq 1}(1 - p)^i\Big(\alpha\sum_{j = 1}^ij(1 - \alpha)^{j - 1}h_2\Big(\frac{1}{j}\Big) \notag \\
&\qquad\qquad\qquad\qquad\qquad\qquad\qquad + i(1 - \alpha)^ih_2\Big(\frac{1}{i}\Big)\Big)\Big]. \notag
\end{align}
However, both $L_2^{\mcl{M}1}$ and $\mathfrak{L}_1^{\mcl{M}1}$ turn out to be better than their SIR counterparts by less than $1\%$.

\subsection{Information Rates for the Sticky Channel}
\begin{figure}[!ht]
\centering
\figselect{\include{bdc}}{\includegraphics{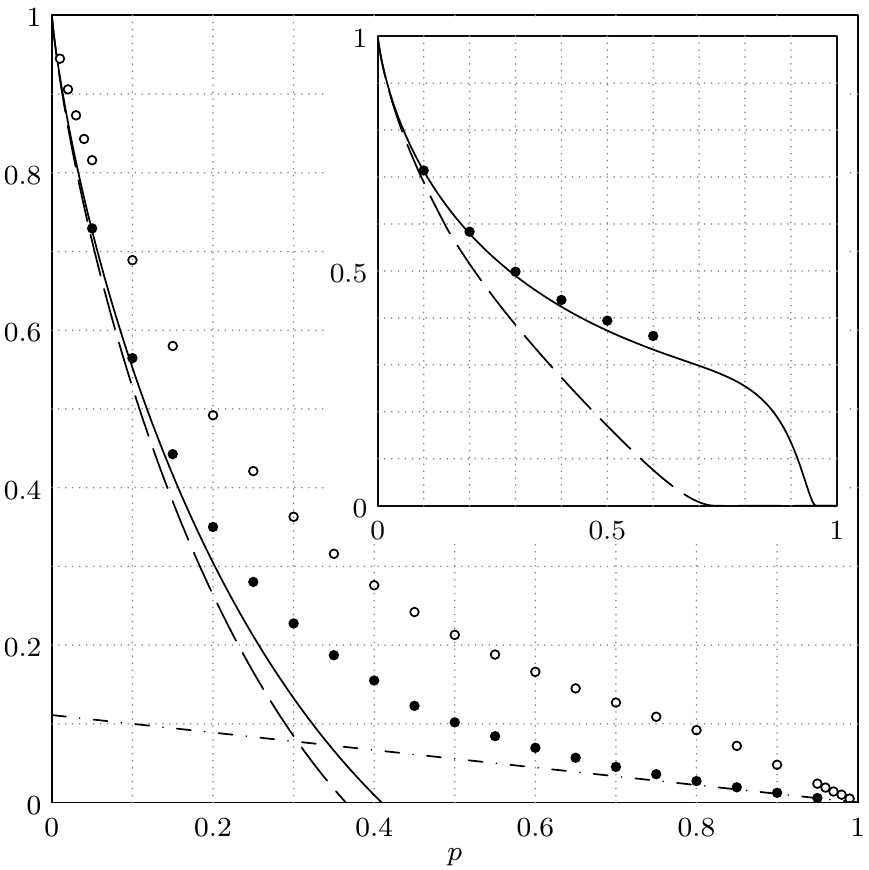}}
\caption{Bounds on the capacity for the binary deletion channel. $L_2^{\mrm{iud}}$ in \eqref{eq_l2sir} is shown as the long-dashed line and $\mathfrak{L}_1^{\mrm{iud}}$ in \eqref{eq_l1sir} (with the $O(p^2)$ term dropped) as the solid line. The best known numerical lower \cite{kir_10_tit_ddccap} and upper bounds \cite{fer_10_tit_bdccap} are shown as black and white circles respectively. The best known lower bound as $p$ approaches $1$ \cite{mit_06_tit_lbcapdc} is shown as the dash-dotted line. The inset plots the bound \eqref{eq_stilb3} as the long-dashed line for the sticky channel, and the Markov-$1$ rate \eqref{eq_stim1} as the solid line. The lower bounds from \cite{mit_08_tit_sticky} are shown as black circles.}
\label{fig_capbounds}
\end{figure}
The analysis for the sticky channel is very similar to that for the BDC in the previous subsection. Since $\lim_{n \rightarrow \infty} \frac{1}{n}I(Y_{[N_n]}; Z_{[N_n]}) \neq 0$ when the input is i.u.d., we bound the $\mcl{H}(\mcl{Z} | \mcl{Y})$ term (see Equation \eqref{eq_ixy}) differently. In this case, we obtain
\begin{align}
C 
&\geq \sup_{\mcl{X}_{\mcl{M}}}\Big[\sup_{i \geq 1}\Big(\mcl{H}(\mcl{X}) + \frac{H(Z_1 | Z_i, \mcl{X}, \mcl{Y}) - H(Z_1 | Y_1)}{1 - p}\Big)\Big] \notag 
\end{align}
\begin{align}
&\geq \sup_{\alpha}\Big[h_2(\alpha) + \sup_{i \geq 1}\Big(ip(1 - p)^{i - 2}H(Z_1 | Z_i = 1, \mcl{X}, \mcl{Y})\Big) \notag \\
&\quad - \frac{p + (1 - \alpha)(1 - p)}{1 - p}h_2\Big(\frac{p}{p + (1 - \alpha)(1 - p)}\Big)\Big], \label{eq_stilb3}
\end{align}
where
\begin{align}
H(Z_1 | Z_i = 1, \mcl{X}, \mcl{Y}) &= \frac{1}{i}\sum_{j = 1}^{i - 1}(j + 1)h_2\Big(\frac{1}{j + 1}\Big)(1 - \alpha)^j\alpha \notag \\
&\qquad\qquad+ h_2\Big(\frac{1}{i}\Big)(1 - \alpha)^i. \notag
\end{align}
For $\alpha = \frac{1}{2}$, we get $C \geq 1 + p\log_2p + dp - O(p^2)$ where $d = \log_2(\frac{2}{e}) + \frac{1}{2}\sum_{j \geq 1}\frac{j}{2^j}\log_2j \approx 0.845836$. As was the case for the BDC, we expect this to be a tight bound for the capacity for small $p$. In fact, for the sticky channel, we can exactly characterize the maximum rate achievable by a first-order Markov process $C^{\mcl{M}1}$ as
\begin{align}
&C^{\mcl{M}1} = \sup_\alpha \Big[h_2(\alpha) \notag \\
&\qquad + \alpha\sum_{r \geq 1}\Big((1 - \alpha)\frac{1 - p}{p}\Big)^r\Big(\sum_{s \geq r} {s\choose r}p^sh_2(\frac{r}{s})\Big) \notag \\
&\qquad - \frac{p + (1 - \alpha)(1 - p)}{1 - p}h_2\Big(\frac{p}{p + (1 - \alpha)(1 - p)}\Big)\Big]. \label{eq_stim1}
\end{align}
Figure \ref{fig_capbounds} plots all the bounds obtained for BDC and sticky channel.

We note that the Markov-$1$ rate \eqref{eq_stim1} is larger than $1 - p$ for a range of $p$ values. This disproves the conjecture that the capacity is convex in $p$ for the sticky channel, unlike what is expected for the BDC \cite{dal_10_arx_capdelub}.

\section{Conclusions} \label{sec_conc}
The model presented here provides a unified framework to handle a broad class of channels with synchronization errors over any finite alphabet. For channels with only deletions or only duplications, we 
obtain analytical lower bounds on the capacity, including some bounds that are expected to be tight for small deletion or duplication probabilities. More generally, the model has an immediate factor-graph interpretation, and this could potentially be used to explore reliable coding schemes. Moreover, it could facilitate the exploration of some fundamental theoretical questions, e.g., establishing a coding theorem for synchronization error channels with memory. A more detailed treatment of some of these questions, results for the BDC and sticky channel, and generalizations to other channels of interest is in preparation \cite{iye_11_tit_sec}.

\section*{Acknowledgment}
The work of A. R. Iyengar is supported by the Center for Magnetic Recording Research and the National Science Foundation under the Grant CCF-$0829865$.

\twobibs{
\bibliographystyle{IEEEtran}
\bibliography{/disks/work/ucsd/Bib/mybib}
}
{
}

\end{document}